\documentclass[12pt]{article}
\usepackage{amsmath}
\usepackage{epsf}
\usepackage{epsfig}
\usepackage{here}
\usepackage{amssymb}
\usepackage{citesort}
\usepackage{graphicx}
\usepackage{latexsym}
\newcommand{\be}{\begin{equation}}
\newcommand{\ee}{\end{equation}}
\newcommand{\bear}{\begin{eqnarray}}
\newcommand{\ear}{\end{eqnarray}}

\newsavebox{\LSIM}
\sbox{\LSIM}{\raisebox{-1ex}{$\ \stackrel{\textstyle<}{\sim}\ $}}
\newcommand{\lsim}{\usebox{\LSIM}}
\newsavebox{\GSIM}
\sbox{\GSIM}{\raisebox{-1ex}{$\ \stackrel{\textstyle>}{\sim}\ $}}

\begin{document}
\begin{titlepage}
\begin{flushright}
BA-02-16\\
DESY 02-045\\
hep-ph/0205327
\end{flushright}
$\mbox{ }$
\vspace{.1cm}
\begin{center}
\vspace{.5cm}
{\bf\Large Majorana Neutrinos}\\[.3cm]
{\bf\Large in a Warped 5D Standard Model}\\
\vspace{1cm}
Stephan J. Huber$^{a,}$\footnote{stephan.huber@desy.de}
and 
Qaisar Shafi$^{b,}$\footnote{shafi@bartol.udel.edu} \\ 
 
\vspace{1cm} {\em  
$^a$Deutsches Elektronen-Synchrotron DESY, Hamburg, Germany}\\[.2cm] 
{\em $^b$Bartol Research Institute, University of Delaware, Newark, USA} 

\end{center}
\bigskip\noindent
\vspace{1.cm}
\begin{abstract}
We consider neutrino oscillations and neutrinoless double beta decay
in a five dimensional standard model with warped geometry. 
Although the see-saw mechanism in its simplest form cannot be implemented
because of the warped
geometry, the bulk standard model neutrinos can acquire the desired (Majorana)
masses from dimension five interactions. We
discuss how large mixings can arise, why the large mixing angle MSW solution
for solar neutrinos is favored, and provide estimates for the mixing angle
$U_{e3}$. Implications for neutrinoless double beta decay are also discussed.
\end{abstract}
\end{titlepage}
\section{Introduction}
Extra dimensions are an interesting possibility for physics beyond
the standard model (SM). They offer an explanation for the weakness
of gravity by the large volume of the compactified space \cite{ADD},
or by localization of the graviton at a space-time boundary \cite{RS,G}.  
Gravity becomes strong at the TeV scale, and the large hierarchy
between the scale of gravity and the weak scale is eliminated.

We take the fifth dimension to be an $S_1/Z_2$ orbifold with 
a negative bulk cosmological constant, bordered by two 3-branes 
with opposite tensions and separated by distance $R$.  Einstein's
equations are satisfied by the non-factorizable metric  \cite{RS}
\begin{equation} \label{met}
ds^2=e^{-2\sigma(y)}\eta_{\mu\nu}dx^{\mu}dx^{\nu}+dy^2, ~~~~\sigma(y)=k|y|
\end{equation}
which describes a slice of AdS$_5$. The 4-dimensional metric is 
$\eta_{\mu\nu}={\rm diag}(-1,1,1,1)$, $k$ is the AdS curvature 
related to the bulk cosmological constant and brane tensions, 
and $y$ denotes the fifth coordinate. The AdS curvature and the 
5d Planck mass $M_5$ are both assumed to be of order $M_{\rm Pl}\sim10^{19}$ GeV.
The AdS warp factor $\Omega=e^{-\pi k y}$
generates an exponential hierarchy of energy scales.
If the brane separation is $kR\simeq 11$, the scale at
the negative tension brane, located at $y=\pi R$, is of TeV-size, 
while the scale at the brane at $y=0$ is of order $M_{\rm Pl}$. 
At the TeV-brane gravity is weak because the zero mode
corresponding to the 4D graviton is localized at the
positive tension brane (Planck-brane). 

Models with localized gravity open up attractive possibilities for flavor 
physics. If the SM fermions reside in the 5-dimensional bulk, the 
hierarchy of quark and lepton masses can be interpreted in a
geometrical way \cite{GP,HS2}. Different flavors are localized at 
different positions in the extra dimension or, more precisely, 
have different wave functions. 
The fermion masses are in direct proportion to the overlap of their wave
functions with the Higgs field \cite{AS}.
Moreover, bulk fermions reduce
the impact of non-renormalizable operators which, for instance,
induce rapid proton decay and  large neutrino masses \cite{GP,HS2}.

In the proposed framework
it is natural to take the SM gauge bosons as bulk fields as well 
to maintain 5D gauge invariance. The Higgs field has to be localized
at the TeV-brane. Otherwise the gauge hierarchy problem
reappears \cite{CHNOY,HS}. Comparison with electroweak
data, in particular with the weak mixing angle and gauge boson
masses, requires the Kaluza-Klein (KK) excitations of SM particles
to be heavier than about 10 TeV \cite{HS,HLS}. If the fermions
were confined to the TeV-brane, the KK scale would be even
more constrained \cite{HS,CET}.\footnote{Recently it has been 
advocated that the third generation of quarks should be confined
to the TeV-brane to prevent large contributions to the electroweak 
$\rho$ parameter \cite{HPR}. However, no such conclusion can 
be drawn for the third generation of leptons.}   

In refs.~\cite{GN,HS3} it was demonstrated that small Dirac 
neutrino masses can be generated by adding sterile neutrinos 
in the bulk. Both the solar and the atmospheric neutrino 
anomalies can be accommodated. 

In this letter we investigate the alternative possibility of 
Majorana neutrinos. Because of the warped geometry, 
the usual see-saw mechanism cannot be implemented 
in a straightforward way. Namely, under the assumption 
that the large 5D Majorana mass is $y$ independent, the 
lowest state of the singlet neutrino, it turns out, is localized 
close to the TeV-brane where it acquires a typical KK scale 
mass ($\sim$ a few TeV) from the warped geometry. 
Fortunately, in contrast to the SM, the warped
5D SM admits larger dimension five Majorana masses and,  
with a judicious choice of parameters these can be 
employed for resolving the solar and atmospheric neutrino 
anomalies. Large mixings arise if
the SM neutrinos are located at similar positions in the
extra dimension. We demonstrate that the large mixing
angle MSW is the favored solution to the solar neutrino
anomaly, while $U_{e3}$ can be kept below the observational
bound. We also discuss possible experimental signatures to 
distinguish between the scenarios of Dirac and Majorana
neutrinos, such as neutrinoless double beta decay and
$\mu\rightarrow e\gamma$. We also comment 
on the issue of proton decay.

\section{Fermions in a Warped Background}
To fix the notation let us briefly summarize the properties
of fermions in a warped geometrical background. Since the
5D theory is non-chiral, every Weyl fermion of the SM has
to be associated with a Dirac fermion in the bulk. The number
of fermionic degrees of freedom is doubled. Chirality in the 4D
low energy effective theory is restored by the orbifold 
boundary conditions.

The Dirac equation for a fermion in curved space-time reads
\begin{equation}
E_a^M\gamma^a(\partial_M+\omega_M)\Psi+m_{\Psi}\Psi=0,
\end{equation}
where $E_a^M$ is the f\"unfbein, $\gamma^a=(\gamma^{\mu},\gamma^5)$ 
are the Dirac matrices in flat space, 
\begin{equation}
\omega_M=\left(\frac{1}{2}e^{-\sigma}\sigma'\gamma_5\gamma_{\mu},0\right)
\end{equation}
 is the spin connection, and $\sigma'=d\sigma/dy$.
The index $M$ refers to objects in 5D curved space,
the index $a$ to those in tangent space. 
Fermions have two possible transformation properties 
under the $Z_2$ orbifold symmetry,
$\Psi(-y)_{\pm}=\pm \gamma_5 \Psi(y)_{\pm}$. Thus, $\bar\Psi_{\pm}\Psi_{\pm}$ 
is odd under $Z_2$, and the Dirac mass parameter, which is also odd, 
can be parametrized as $m_{\Psi}=c\sigma'$. The Dirac mass should 
therefore originate from the coupling to a $Z_2$ odd scalar field 
which acquires a vev. 
On the other hand, $\bar\Psi_{\pm}\Psi_{\mp}$ is even.
Using the metric (\ref{met}) one obtains for the left- and right-handed components
of the Dirac spinor \cite{GN,GP}
\begin{equation} \label{eom}
[e^{2\sigma}\partial_{\mu}\partial^{\mu}+\partial_5^2-\sigma'\partial_5-M^2]
        e^{-2\sigma}\Psi_{L,R}=0,
\end{equation} 
where $M^2=c(c\pm1)k^2\mp c\sigma''$ and  $\Psi_{L,R}=\pm\gamma_5\Psi_{L,R}$.

Decomposing the 5d fields as 
\begin{equation}
\Psi(x^{\mu},y)=\frac{1}{\sqrt{2\pi R}}\sum_{n=0}^{\infty}\Psi^{(n)}(x^{\mu})f_n(y),
\end{equation}
eq.~(\ref{eom}) admits a zero mode solution \cite{GN,GP}
\begin{equation} \label{f0}
f_0(y)=\frac{e^{(2-c)\sigma}}{N_0},
\end{equation}
and a tower of KK excited states
\begin{equation} \label{f1}
f_n(y)=\frac{e^{5\sigma/2}}{N_n}\left[J_{\alpha}(\frac{m_n}{k}e^{\sigma})+
                 b_{\alpha}(m_n)Y_{\alpha}(\frac{m_n}{k}e^{\sigma})\right].
\end{equation}
The order of the Bessel functions is $\alpha=|c\pm 1/2|$ for $\Psi_{L,R}$.
The spectrum of KK masses $m_n$ and the coefficients $b_{\alpha}$
are determined by the boundary conditions of the wave functions at the 
branes.
The normalization constants follow from
\begin{equation}
\frac{1}{2\pi R}\int^{\pi R}_{-\pi R}dy~e^{-3\sigma}f_m(y)f_n(y)=\delta_{mn}.
\end{equation}

Because of the orbifold boundary conditions, the zero mode of 
$\Psi_+$ $(\Psi_-)$ is a left-handed (right-handed) Weyl spinor.  
For $c>1/2$ $(c<1/2)$ the zero mode of the fermion is localized 
near the boundary at $y=0$ $(y=\pi R)$, i.e.~at the Planck-  (TeV-) brane.
The KK states are always localized at the TeV-brane.

\section{The See-Saw Mechanism and Warped Geometry}
In the see-saw mechanism lepton number is broken
by the Majorana mass of heavy right-handed neutrinos,
${\cal L}_M=M_NNN$. Together with a Dirac mass term
${\cal L}_D=\lambda_N\nu_LN  H $, where $H$ denotes the SM Higgs
field,  a Majorana
mass for the SM neutrinos $\nu_L$ is generated,
\begin{equation} \label{ss}
M_{\nu}=\frac{\lambda_N^2 \langle H \rangle^2}{M_N}.
\end{equation}
Taking $M_{\nu}\sim 50$meV (of the order of the
atmospheric $\Delta m^2$), one finds 
$M_N\sim \lambda_N^2\cdot6\times10^{14}$ GeV.
For $0.01\lsim\lambda_N\lsim1$ this points to an intermediate
scale for the right-handed Majorana mass. 

In theories with TeV-scale gravity $M_N$ is naturally bounded
to be below a few TeV. Therefore, very small
Yukawa couplings $\lambda_N\lsim10^{-6}$ are required
to generate sub-eV neutrino masses. Keeping in mind the
large hierarchy of Yukawa couplings for quarks and 
charged leptons, this result may  not seem ``completely unreasonable''.
Nonetheless, a deeper understanding of the smallness of
neutrino masses is missing.

If the SM and right-handed neutrinos reside in the bulk,
the 5D Yukawa coupling $\lambda_N^{(5)}$ is somewhat
less constrained than the 4D coupling we were concerned 
with in the previous paragraph. The Dirac mass which enters
eq.~(\ref{ss}) depends on the overlap of the neutrino
zero modes and the Higgs
\begin{equation} \label{3.2}
m_D\equiv\lambda_N\langle H\rangle=\int_{-\pi R}^{\pi R}\frac{dy}{2\pi R}
\lambda^{(5)}_{N}e^{-4\sigma}H(y) f_{0}^{(\nu)}(y)f_{0}^{(N)}(y).
\end{equation}
We assume that the Higgs profile has an exponential form
which peaks at the TeV-brane,
$H(y)=H_0e^{4k(|y|-\pi R)}.$
This shape can be motivated by the equation of motion
of a bulk scalar field \cite{GW}. Numerically it is equivalent
to a delta function-like profile. 
Using the known mass of the W-boson, we can fix the amplitude 
$H_0\sim M_{\rm Pl}$ in terms of the 5D weak gauge coupling.

The zero mode of the SM neutrino has an exponential
shape (\ref{f0}) which for $c_{\nu}>1/2$ has only a small
overlap with the Higgs at the TeV-brane. Sticking to the
idea that all input parameters in the 5D theory should be 
of order unity in natural units, we assume that in the bulk 
lepton number is broken at $M_N^{(5)}\sim M_{\rm Pl}$. 
Because of this large Majorana mass, the wave 
function of the lowest state of the right-handed neutrino
is modified. Rather than acquiring a mass of order $M_{\rm Pl}$,
this state gets localized at the TeV-brane where the
warped geometry induces a typical KK mass for it, very similar
to excited fermionic states.  As a result, the overlap of the 
right-handed neutrino and the Higgs is large. These basic properties 
of the right-handed ``zero mode'' are fairly independent of the details of 
the KK reduction in the presence of a large bulk Majorana mass.\footnote{
If one allows for a suitably tuned profile of the Majorana mass in the bulk, the
lowest state of the right-handed neutrino can be peaked around the
Planck-brane. Its mass is still found to be bounded by the KK scale.}
They also do not depend on a possible 5D Dirac mass $c_N$ for
the right-handed neutrino.

\begin{figure}[t] 
\begin{picture}(100,160)
\put(95,-10){\epsfxsize7cm \epsffile{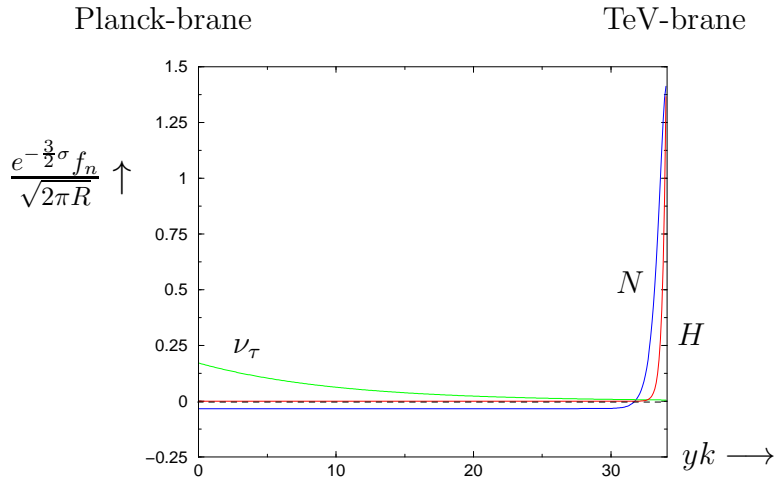}}
\put(45,100){{\large $\frac{e^{-\frac{3}{2}\sigma}f_n}{\sqrt{2\pi R}}\uparrow$}} 
\put(130,37){{$\nu_{\tau}$}} 
\put(275,60){{$N$}} 
\put(298,40){{$H$}} 
\put(300,-5){$yk\longrightarrow$}
\put(270,160){{TeV-brane}} 
\put(70,160){{Planck-brane}} 
\end{picture} 
\caption{Wave functions of the $\nu_{\tau}$ and $N$ zero modes, together
with a sketch of the Higgs profile.
}
\label{f_0}
\end{figure}

We now estimate the required value of $\lambda_N^{(5)}$
to generate a 50 meV neutrino mass. As explained above, the
Majorana mass of the zero mode $M_N$ is close to the KK scale 
$M_{KK}$. From comparison with electroweak precision data, 
$M_{KK}$ cannot be smaller than about 10 TeV \cite{HLS}. 
By localizing $f_{0}^{(\nu)}$ close to the Planck-brane 
($c_{\nu}\gg1/2$), $m_D$ can be made very small. However,
at the same time, the mass of the charged lepton, residing
in an SU(2) multiplet with the neutrino, can become very small 
as well \cite{GP,HS2}. Since the tau is the heaviest lepton,
it has to reside closest to the TeV-brane, and induces the
largest $m_D$. Assuming 5D lepton Yukawa couplings
of order unity (i.e.~$\lambda_L^{(5)}\sim g^{(5)}$, the 5D weak gauge 
coupling),  the left-handed tau has to obey $c\lsim 0.57$ (when the
right-handed tau is delocalized, $c=$1/2) \cite{HS2}.  If
we take $M_N=10$ TeV, $\lambda_N^{(5)}=g^{(5)}$  
and $c_{\tau}= 0.57$, we find $m_D=7.9$ GeV. A neutrino
mass of 50 meV requires $\lambda_N^{(5)}/g^{(5)}\sim9\times10^{-5}$.
In accordance with the discussion in the previous paragraph,
we have approximated the right-handed zero mode by the
wave function of the first KK state of a fermion (\ref{f1}) without
Majorana mass and $c_N=1/2$. In fig.~\ref{f_0} we present 
the associated wave functions of the $\nu_{\tau}$ and $N$ 
zero modes, together with the Higgs profile.
The bound on the Yukawa coupling only mildly depends
on $c_N$, increasing it from 1/2 to 1 would result in    
$\lambda_N^{(5)}/g^{(5)}\sim7\times10^{-5}$. The error
arising from our approximation of the right-handed 
neutrino wave function is of the same level.

The bound on
$\lambda_N^{(5)}$ can be somewhat relaxed if we assume that
the right-handed tau is localized at the TeV-brane (let's say
with $c=-1$). Then we can move the left-handed tau and the
tau neutrino closer to the Planck brane, $c\sim 0.63$, and
find  $\lambda_N^{(5)}/g^{(5)}\sim6\times10^{-4}$.  
However, having the right-handed leptons localized towards
the TeV-brane ($c<1/2$) induces large deviations in the 
electroweak observables \cite{HLS}. The KK scale is 
required to be much higher then 10 TeV, and the 
gauge hierarchy problem gets reintroduced, albeit in a ``mild'' form.
In the case
of the muon the constraints on $\lambda_N^{(5)}$ are an order of 
magnitude weaker.

If one relies on supersymmetry to solve the gauge hierarchy problem,
the Higgs fields can reside in the bulk. Then a different and more 
standard version of the see saw mechanism may be implemented: 
The right-handed neutrinos, having an intermediate mass, are confined 
to the Planck-brane. This translates into an intermediate see saw 
scale in the effective 4D theory. Since the Higgs is spread homogeneously
in the bulk, there are no small overlaps anymore. Thus, an
explanation of the fermion mass hierarchy is lost.  An alternative way
of implementing the see-saw mechanism is to assume a $y$ dependent
singlet Majorana mass. 

In conclusion, having the right-handed and sterile neutrinos
in the bulk reduces the fine-tuning of the neutrino Yukawa 
coupling in the see-saw approach to some extent. However, 
a small number of order $10^{-4}$ is still needed to 
generate sub-eV neutrino masses.

\section{Dimension Five Neutrino Masses}
In the 4D SM one does not have to worry about possible higher-dimensional
operators, since they are safely suppressed by powers of the huge 
Planck mass. Models with a low scale of gravity are very different
in this respect. Naively one expects exotic non-renormalizable 
interactions suppressed by only a few TeV mass scale, unless
they are 
forbidden by symmetries or multiplied by tiny coupling constants.   
In the warped SM the suppression scale for non-renormalizable
operators can be anywhere between a few TeV and the Planck scale,
depending on the localization of the fermions \cite{GP,HS2}.

As previously discussed in refs.~\cite{GP,HS2}, 
Majorana masses for left-handed neutrinos are generated by the 
dimension-five operator
\begin{equation} \label{nu3}
\int d^4x\int dy \sqrt{-g}\frac{l_{ij}}{M_5^2}H^2\Psi_{iL}C\Psi_{jL}
\equiv \int d^4x ~M^{(\nu)}_{ ij}\Psi_{iL}^{(0)}C\Psi_{jL}^{(0)},
\end{equation}
where $l_{ij}$ are dimensionless couplings constants and 
$C$ is the charge conjugation operator. The neutrino mass matrix
reads
\begin{equation} \label{nu}
M^{(\nu)}_{ ij}=\int_{-\pi R}^{\pi R}\frac{dy}{2\pi R}\frac{l_{ij}}{M_5^2}e^{-4\sigma(y)}
H^2(y)f_{0i}^{(\nu)}(y) f_{0j}^{(\nu)}(y).
\end{equation}
Again the most stringent constraint on $l$ comes from the tau neutrino.
Assuming the Yukawa coupling of the tau to be $\lambda_{\tau}^{(5)}\approx g^{(5)}$
and $c(\tau_R)=1/2$, we found  $c(\tau_L)=0.567$. From eq.~(\ref{nu}) 
we obtain a tau neutrino mass of $m_{\nu,\tau}=l\cdot 33$ MeV \cite{HS2}.
Bringing the neutrino mass down to 50 meV requires a very small 
value of $l\sim10^{-9}$.  Such a small coupling could originate from
non-perturbative effects of gravity, especially if there is an extra 
dimension somewhat larger than $M_{\rm PL}$ \cite{KKLLS}. 
However, it would be attractive to relax the bound on
$l$ within the framework of the effective theory we 
have considered so far.  

We can reduce the neutrino masses by shifting the left-handed
leptons closer to the Planck-brane. To maintain the masses of
the charged leptons, we either have to move the right-handed
leptons closer to the TeV-brane or increase the 5D Yukawa
couplings of the leptons $\lambda^{(5)}$ .  The first possibility is clearly disfavored
by the electroweak precision data, as explained in the previous
section. Keeping the positions of the right-handed leptons fixed at $c=1/2$,
the neutrino masses are roughly proportional to $1/(\lambda^{(5)})^2$ 
\cite{HS2}. If we restrict ourselves to  $\lambda_{\tau}^{(5)}=10g^{(5)}$,
in order not to induce a large hierarchy in the 5D couplings, we
find  $c(\tau_L)=0.647$ and $m_{\nu,\tau}=l\cdot 320$ keV. 

The neutrino masses can be further decreased if we take the
AdS curvature $k$ to be smaller than the 5D Planck mass.
Approximately, the relation $m_{\nu}\propto (k/M_5)^2$ holds \cite{HS2}. 
The assumption $k<M_5$ is anyway necessary to derive
the warped metric (\ref{met}) from Einstein equations, neglecting
terms with higher derivatives \cite{RS}. In discussions of the collider
phenomenology of the warped model, the parameter range 
$M_5/100<k<M_5$ has been considered \cite{DHR}. Taking now the
the very favorable case $\lambda_{\tau}^{(5)}=10g^{(5)}$ and
$k/M_5=0.01$, we obtain  $m_{\nu,\tau}=l\cdot 25$ eV. A neutrino
mass of 50 meV can therefore be generated by $l=0.002$, which
is only a moderately small number. However, one has to keep in mind
that for $k/M_5=0.01$ the radius $kR=8.63$ is already of order
$10^3$ in units of the fundamental scale $M_5$.

Thus, the Majorana neutrino masses induced by the 
dimension five operator (\ref{nu3}) are quite large if we take
the model parameters to be strictly of order unity. In ref.~\cite{HS2}
we therefore imposed lepton number to eliminate (\ref{nu3})
completely.
Here we assume that one or a combination of the mechanisms
discussed above is at work to bring down the neutrino masses
to values which are consistent with experimental observations.
In the following we investigate the consequences for neutrino 
mixing as well as for neutrinoless double beta decay.

\section{Neutrino Mixings}
In ref.~\cite{HS2} it was found that the dimension five
neutrino masses (\ref{nu}) 
lead to small mixing angles for the neutrinos. This result 
arose from the different locations  
of the left-handed lepton flavors in the extra dimension 
in order to explain the hierarchy of charged lepton masses.  
In the following we demonstrate that large mixing angles
naturally arise if the left-handed leptons have the same location
in the extra dimension. This result is similar to the case of
Dirac masses for neutrinos studied in ref.~\cite{HS3}. 
We will also show that separating the electron neutrino somewhat
from the muon and tau neutrinos helps to keep $U_{e3}$ 
sufficiently small.

To obtain the neutrino masses and mixings we diagonalize
the mass matrix  (\ref{nu}) with a unitary matrix 
$M^{(\nu)}=U_{\nu}M_{\rm diag}^{(\nu)}U^T_{\nu}$. 
The physical neutrino mixings
\begin{equation}
U=U_l^{\dagger}U_{\nu}
\end{equation}
also depend on the rotations of the left-handed charged 
leptons $U_l$. The charged lepton masses $M^{(l)}$ arise from their
couplings to the Higgs field in the same way as the Dirac neutrino
 masses of eq.~(\ref{3.2}),
\begin{equation} \label{3.2a}
M^{(l)}_{ij}=\int_{-\pi R}^{\pi R}\frac{dy}{2\pi R}
\lambda^{(5)}_{N}e^{-4\sigma}H(y) f_{0i}^{(l)}(y)f_{j}^{(e)}(y),
\end{equation}
where we have inserted the charged lepton wave functions and Yukawa
couplings $\lambda_l^{(5)}$ \cite{GP,HS2}. 
We determine $U_l$ by diagonalizing $M^{(l)}(M^{(l)})^{\dagger}$.
In general, the neutrino and charged lepton rotations are
expected to show the same pattern on average since their positions
in the extra dimension are tied together by the SU(2) gauge
symmetry of the SM. 
Including CP violation, the mixing matrix $U$ can be written as 
\begin{equation}
U=  \left(\begin{array}{ccc} 
c_2c_3 & c_2s_3 & s_2e^{-i\delta}  \\[.1cm] 
-c_1s_3-s_2s_1c_3e^{i\delta} & c_1c_3-s_2s_1s_3e^{i\delta} & c_2s_1 \\[.1cm]  
s_1s_3-s_2c_1c_3e^{i\delta} &  -s_1c_3-s_2c_1s_3e^{i\delta} & c_2c_1  
\end{array}\right),
\end{equation}  
where $s_i$ and $c_i$ are the sines and cosines of $\theta_i$. Like in the
CKM matrix, there is a single complex phase $\delta$ which induces
CP violation in the lepton sector.  The neutrino mass matrix contains
two additional Majorana phases which, however, do not show up in
the mixing matrix $U$.

The atmospheric neutrino data imply \cite{SK}
\begin{equation} \label{atm}
1\cdot10^{-3}{\rm eV}^2 <\Delta m^2_{\rm atm}<5\cdot10^{-3}{\rm eV}^2,
\quad \sin^22\theta_1>0.85.
\end{equation}
There are several solutions to the solar neutrino anomaly \cite{BGP}
\begin{equation} \label{sol}
\begin{array}{ccc} 
 &  \Delta m^2_{\rm sol}~[{\rm eV}^2]\quad  &  \sin^22\theta_3\quad\\[.2cm]
{\rm LMA}\quad &2\cdot10^{-5}- 4\cdot10^{-4} & 0.3-0.93 \\[.2cm]
{\rm SMA}\quad &\quad4\cdot10^{-6}- 9\cdot10^{-6}\quad  & \quad0.0008-0.008\quad  \\[.2cm]
{\rm LOW}\quad &6\cdot10^{-8}- 2\cdot10^{-7} & 0.89-1 \\[.2cm]
{\rm VAC}\quad &\sim10^{-10} & 0.7-0.95 
\end{array}
\end{equation}
The SMA solution gives only a poor fit to the data. The CHOOZ reactor
experiment constrains $|U_{e3}|^2\equiv s_2^2$ to be at most a few 
percent \cite{CHOOZ}. Nothing is known about the CP violating phase $\delta$. 

Our solution to the neutrino puzzle follows the ``neutrino mass anarchy'' 
models \cite{HMW00}. Mass matrices with randomly chosen entries
of order unity have a large probability to fit the neutrino data. Thus,
we parametrize the coefficients in (\ref{nu}) by
\begin{equation}
l_{ij}\equiv \Lambda\cdot \tilde l_{ij}.
\end{equation}
In contrast to the 4D realization of ``neutrino mass anarchy'', in
our scenario the smallness of the neutrino masses is not attributed
to a very small overall magnitude $\Lambda$ of the couplings. Rather,
the neutrinos are localized close to the Planck-brane, where
the dimension five operator (\ref{nu3}) is suppressed. The supposedly 
fundamental theory responsible for the effective interaction 
(\ref{nu3}) is represented through the order unity coefficients $\tilde l_{ij}$.
To incorporate the charged lepton rotations we also generate 
charged lepton mass matrices with random Yukawa couplings of
order unity. We choose suitable lepton locations to reproduce the
measured lepton masses. 

We consider the very favorable case of $\lambda_{\tau}^{(5)}=10g^{(5)}$ and
$k/M_5=0.01$. We take the position of the right-handed tau
to be $c_{\tau,R}=1/2$, which means its wave function is delocalized
in the extra dimension. To reproduce the tau mass we find the
position of the left-handed  tau to be $c_{\tau,L}=0.72$. Since different
positions for the left-handed leptons tend to produce small
neutrino mixing angles \cite{HS2,HS3}, we take 
\begin{equation}
c_{e,L}=c_{\mu,L}=c_{\tau,L}=0.72.
\end{equation}
To accommodate the muon and electron masses we use
$c_{\mu,R}=0.64$ and $c_{e,R}=0.85$. 
Taking the order unity coefficients to be homogeneously 
distributed $1/2<|\tilde l_{ij}|<2$ with random phases from 0 to $2\pi$, we find
the most favorable value for the overall scale to be 
$\Lambda=1.9\cdot10^{-3}$. We randomly generate 
parameter sets for $\tilde l_{ij}$, calculate the neutrino
mass matrix from (\ref{nu}) and compute the neutrino
masses and mixings. For the charged lepton rotations we take
random Yukawa couplings  
$10g^{(5)}/\sqrt{2}<|[\lambda_l^{(5)}]_{ij}|<10g^{(5)}\cdot \sqrt{2}$ 
with phases from 0 to $2\pi$. The charged lepton masses
and left-handed rotations are then calculated from the
mass matrix  eq.~(\ref{3.2a}). We require the computed
lepton masses to agree up to a factor of 3/2 with the measured values,
which holds for about 50\% of the random sets of $\lambda_l^{(5)}$.
These sets are the starting point for the investigation of the neutrino
properties.

\begin{figure}[t] 
\begin{picture}(100,160)
\put(95,-5){\epsfxsize7cm \epsffile{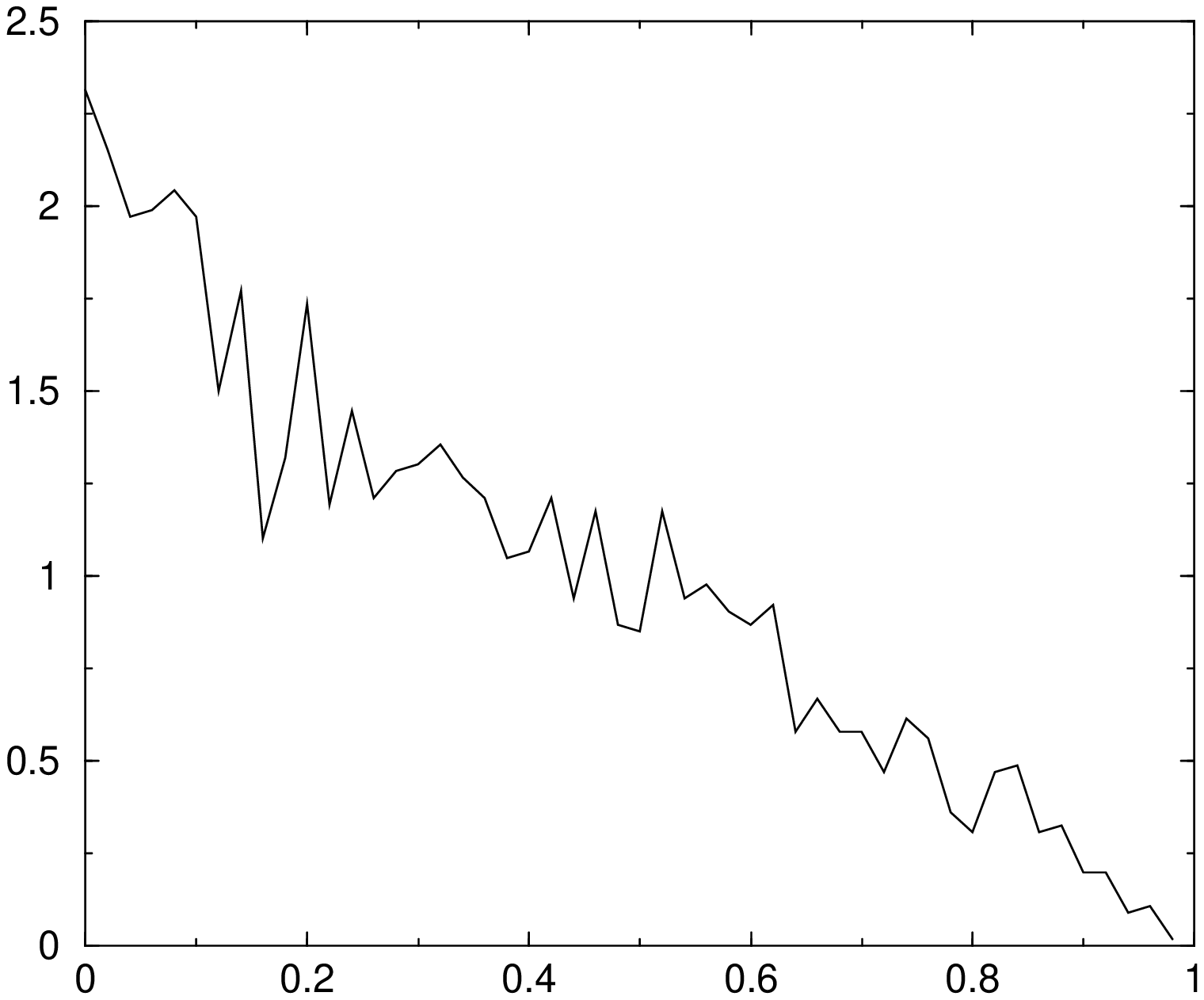}}
\put(60,100){{$\#\uparrow$}} 
\put(305,-5){$|U_{e3}|^2\rightarrow$}
\end{picture} 
\caption{Distribution of $|U_{e3}|^2$ for the case 
$c_{e,L}=c_{\mu,L}=c_{\tau,L}$ and $1/2<|\tilde l_{ij}|<2$.
}
\label{f_1}
\end{figure}

Focusing on the LMA solution of the solar neutrino anomaly,
which turns out to be clearly favored, we find the following 
picture. From our neutrino parameter sets about 70 \% reproduce
$\Delta m^2_{\rm atm}$ (\ref{atm}). Imposing in addition the
constraint from $\Delta m^2_{\rm sol}$ (\ref{sol}), we are
left with about 28 \% of the parameter sets. The solar
and atmospheric mixings angles bring this number down
to about 6 \%, which is still a considerable fraction 
given the number of constraints. The most stringent
constraint turns out to come from the CHOOZ experiment.
Adding the requirement $|U_{e3}|^2<0.05$, the fraction
of viable parameter sets shrinks to about 0.7 \%.
This result is clearly related to $\langle |U_{e3}|^2\rangle =0.22$,
which is considerably above the experimentally favored
value. Here we use the logarithmic average of a quantity $X$
\begin{equation}
\langle X\rangle =\exp\left(\sum_i^N \frac{\ln(X_i)}{N}\right).
\end{equation}
If not stated otherwise, we average over the parameter sets
which reproduce the correct $\Delta m^2$'s. In fig.~\ref{f_1}
we display the distribution of $|U_{e3}|^2$
found in our statistical analysis.
We included only parameter sets that  reproduce the correct $\Delta m^2$'s.
Small values of $|U_{e3}|^2$ are somewhat favored.
The corresponding distributions for the solar and atmospheric
mixing angles are peaked at maximal mixings.
Averaging only over the sets which pass all experimental constraints,
we find $\langle |U_{e3}|^2\rangle =0.022$.
This means that  $|U_{e3}|^2$ is probably close to the experimental bound
and can very likely be tested at future neutrino experiments, 
such as MINOS, which is sensitive down to $|U_{e3}|^2\sim0.0025$ \cite{MINOS}.
Since we typically start with large phases in the neutrino and
charged lepton mass matrices, the complex phase $\delta$ in the mixing 
matrix $U$ is found to be most likely of order unity.
If we ignore the charged lepton rotations, i.e.~take $U=U_{\nu}$, 
the fraction of parameter sets satisfying all constraints rises somewhat
from 0.7 \% to 1.7 \%. The reason is that cancellations amongst the large
mixing angles in $U_l$ and $U_{\nu}$ make it slightly more difficult 
to reproduce large solar and atmospheric mixings.

\begin{table}[t] \centering
\begin{tabular}{|c||c|c|c|} \hline
&  $\Delta m^2_{\rm atm, sol}$ &$+\sin^22\theta_{\rm atm, sol}$ & $+|U_{e3}|^2<0.05$  
\\ \hline 
LMA & 44.4 (28.1) & 5.8 (5.9) & 5.0 (0.7)   \\ \hline
SMA & 1.3 (0.04) & 0.3 ($<0.001$) & 0.3 ($<0.001$)  \\ \hline
LOW & 0.008 ($<0.001$) & 0.002 ($<0.001$)  &  0.002 ($<0.001$)  \\ \hline 
\end{tabular} 
\caption{Probability in percent that a randomly generated set of coefficients
$1/2<|\tilde l_{ij}|<2$ satisfies the constraints from $\Delta m^2_{\rm atm, sol}$
(first column) and $\sin^22\theta_{\rm atm, sol}$ (second column) and
 $|U_{e3}|^2<0.05$  (third column). The results are given for the case
$c_{e,L}>c_{\mu,L}=c_{\tau,L}$ ($c_{e,L}=c_{\mu,L}=c_{\tau,L}$). 
 }
\label{t_L} 
\end{table}

These results depend  only mildly on the interval chosen
for the coefficients $\tilde l_{ij}$. If we narrow the range, let's
say to $0.8<|\tilde l_{ij}|<1.7$, the fit for the  $\Delta m^2$'s,
the solar and atmospheric mixings angles improves remains
unchanged,
while the CHOOZ constraint becomes slightly harder to satisfy. 
We find that now about 0.6 \% of the
parameter sets meet the constraints from eqs.~(\ref{sol}) and (\ref{atm})
and $|U_{e3}|^2<0.05$.

The SMA, LOW and VAC solutions to the solar neutrino anomaly 
can be realized only with severe fine-tuning. The required
small values of $\Delta m^2_{\rm sol}$ are very unlikely to
be produced through accidental cancellations in the neutrino mass matrix.  
For the SMA we find that for the most favorable value of
$\Lambda=1.6\cdot10^{-3}$  and $1/2<|\tilde l_{ij}|<2$, 
only about  0.04 \% of the parameter
sets reproduce the correct $\Delta m^2$'s.  Including the constraints
from the solar and atmospheric mixing angles and the CHOOZ
experiment brings the fraction down to less than $10^{-5}$. 
The LOW and VAC solutions are even more fine-tuned. Under the 
given assumptions the LMA 
case is therefore selected as the only realistic solution to the solar 
neutrino problem. These results, which are summarized in table \ref{t_L},
agree nicely with the 4D model studied in
ref.~\cite{HMW00}. Note that for the LMA solution the results remain
stable if we use real instead of complex mass matrixes. However, the SMA, 
LOW and VAC solutions are more difficult to accommodate with complex
mass matrices since the required cancellations to obtain a small 
 $\Delta m^2_{\rm sol}$ become even more unlikely. For instance, 
with real mass matrices, the fraction of parameter sets which
reproduce the SMA mass squared differences increases from
0.04 \% to 0.7 \%. 

So far the constraint from the CHOOZ experiment is the most 
stringent one. Its inclusion reduces the probability that a parameter 
set realizes the LMA solution from about 6 \% to
less than one percent. With all entries in the neutrino mass matrix being
of similar magnitude, the ensuing mixing angles are typically
large. The fit to the neutrino data  improves considerably if the 
electron neutrino is somewhat separated from the other two
neutrino species. Shifting the electron neutrino closer to the 
Planck-brane induces small elements in the neutrino mass
matrix. As a result, small values of  $\Delta m^2_{\rm sol}$
and $|U_{e3}|^2$ become more probable. The neutrino mass 
matrix (\ref{nu}) 
acquires the following structure
\begin{equation} \label{eps}
M^{(\nu)}\sim\left(\begin{array}{ccc} 
\epsilon^2 & \epsilon & \epsilon  \\[.1cm] 
\epsilon & 1 & 1 \\[.1cm]  
\epsilon & 1 & 1  
\end{array}\right),
\end{equation}  
where $\epsilon\approx f_{0,e}^{(\nu)}(\pi R)/f_{0,\mu}^{(\nu)}(\pi R)
\approx \exp(-(c_{e,L}-c_{\mu,L})\pi kR)$. The charged lepton mass
matrix does not follow the pattern (\ref{eps}) because of the different 
locations of left- and right-handed leptons. Still $U_l$ and $U_{\nu}$

\begin{figure}[t] 
\begin{picture}(100,160)
\put(95,-10){\epsfxsize7cm \epsffile{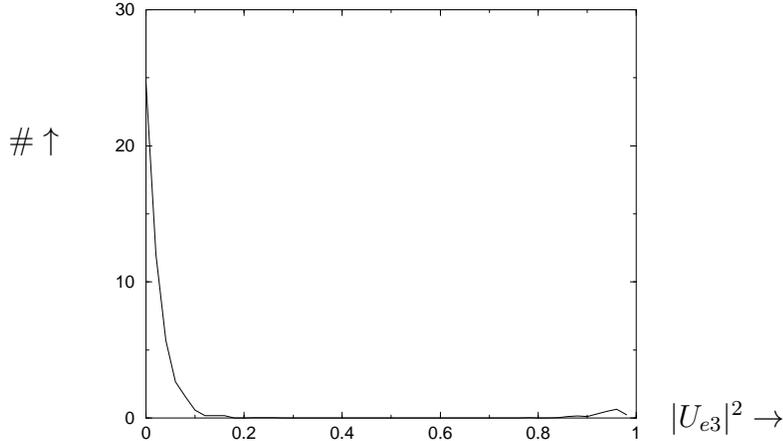}}
\put(55,100){{$\#\uparrow$}} 
\put(305,-5){$|U_{e3}|^2\rightarrow$}
\end{picture} 
\caption{Distribution of $|U_{e3}|^2$ for the case 
$c_{e,L}>c_{\mu,L}=c_{\tau,L}$ and $1/2<|\tilde l_{ij}|<2$.
}
\label{f_2}
\end{figure}

We keep the muon and tau neutrinos at the previous locations 
$c_{\mu,L}=c_{\tau,L}=0.72$. A separation of these two
would make the fit to the atmospheric neutrino data
more difficult since the corresponding mixing is reduced.
(A small separation, e.g.~$|c_{\mu,L}-c_{\tau,L}|\lsim0.02$
could, however, be tolerated.)
For the electron neutrino
the most favorable choice turns out to be $c_{e,L}=0.79$ leading to
$\epsilon=0.15$. 
This value is of the order of  
$\sqrt{m_{\mu}/m_{\tau}}$, and also of the Cabbibo angle. However, in our
model there is no relation to either of these quantities.  The positions
of the right-handed leptons are found to be $c_{e,R}=0.79$,
$c_{\mu,R}=0.62$ and $c_{\tau,R}=0.49$.

Let us again focus on the LMA solution.
Assuming $1/2<|\tilde l_{ij}|<2$ for the order unity coefficients,
we find the best value  $\Lambda=2.6\cdot10^{-3}$. The correct values
of the $\Delta m^2$'s are fitted by 44 \% of the parameter sets.
Taking into account also the constraints from the solar and
atmospheric mixing angles, this fraction shrinks to 5.8 \%.  
This reduction is mostly due to the solar mixing angle which is
suppressed if $\epsilon$ is small. 
Most important, however, the CHOOZ constraint $|U_{e3}|^2<0.05$
is now satisfied almost automatically, and we are finally left with 5.0 \%
of the parameter sets. Compared to the case of $\epsilon=1$, the
probability for a parameter set to satisfy all observational
constraints is enhanced by almost an order of magnitude.
It turns out that including phases in the mass matrices is
very important to obtain this result. With real valued mass 
matrices there tend to be large cancellations between 
the $U_{\nu}$ and $U_l$ contributions to the solar and 
atmospheric mixing angles. Large mixings are then very 
difficult to realize.
If we consider the interval for the coefficients to be $0.8<|\tilde l_{ij}|<1.7$,
the fit becomes slightly better because the solar mixing angle
gets somewhat enhanced. We find that 5.5 \% of the parameter sets
pass all the constraints.

In fig.~\ref{f_2} we display the distribution
of $|U_{e3}|^2$ for the case $1/2<|\tilde l_{ij}|<2$, where we again
include only the parameter sets which reproduce the correct
 $\Delta m^2$'s. As expected, the distribution is now peaked at
small values of $|U_{e3}|^2$. Averaging only over parameter sets
which satisfy all constraints, we find a moderately small value 
$\langle |U_{e3}|^2\rangle=0.010$. As a consequence,
the next generation neutrino experiments still has a good chance
to detect a non-vanishing $|U_{e3}|^2$ \cite{MINOS}.

Taking even smaller values of $\epsilon$, it is possible to
implement the SMA solution to the solar neutrino problem. The most
favorable choice of parameters we find to be $c_{e,L}=0.96$ corresponding 
to  $\epsilon=0.0020$ and $\Lambda=2.6\cdot10^{-3}$. The correct
values of $\Delta m^2$ are reproduced by 1 \% of the sets of
coefficients $1/2<|\tilde l_{ij}|<2$. The constraints from the
solar and atmospheric mixing angles reduce this amount to about 0.3 \%.
The CHOOZ constraint is always satisfied, and we obtain
$\langle |U_{e3}|^2\rangle=2.7\cdot10^{-6}$ which is much too small
to be measurable. 
The LOW solution is very difficult to implement because of the
small solar  $\Delta m^2$. We take $c_{e,L}=0.86$ corresponding 
to  $\epsilon=0.0026$, and $\Lambda=2.4\cdot10^{-3}$. The constraints
from the $\Delta m^2$'s are satisfied by only about 0.01 \% of the parameter sets. 
All the neutrino data are reproduced by a fraction smaller then $10^{-5}$. The 
CHOOZ constraint is again satisfied automatically. Small   $\Delta m^2$'s
could be accommodated by moving the electron (and muon) neutrino
closer to the Planck-brane. Then it becomes however more difficult to
obtain large mixings.
The VAC solution is even more
difficult to realize because of the very small value of $\Delta m^2_{\rm sol}$.
Thus, the LMA solution is
by far the most favored scenario. A collection of our results is given
table \ref{t_L}.

Neutrino mass matrices of the type (\ref{eps}) have previously been
considered in refs.~\cite{V,SY}. There the small quantity 
$\epsilon$ was attributed to a Froggatt-Nielsen mechanism \cite{FN}. Our 
findings  agree very well with the  results of these studies.

\section{Corrections from KK Neutrinos}   
So far we have only considered the neutrino zero modes. But
eq.~(\ref{nu3}) also induces mixings between the zero modes
and the vector-like excited states. In the following we show that this effect
does not modify the conclusions we have reached above.
The general neutrino mass matrix takes the symmetric form
\begin{equation} \label{nu_mass}
{\cal M}_{\nu}= (\nu_L^{(0)},\nu_L^{(1)}, \nu_R^{(1)},\dots) \left(\begin{array}{cccc} 
m^{(0,0)}_L & m^{(0,1)}_L & 0 &\cdots \\[.1cm] 
m^{(1,0)}_L & m^{(1,1)}_L & m_{KK,1}& \cdots \\[.1cm]  
0 &  m_{KK,1} & m^{(1,1)}_R & \cdots \\
\vdots & \vdots & \vdots & \ddots
\end{array}\right)\left(\begin{array}{c}\nu_L^{(0)} \\[.1cm]   \nu_L^{(1)} \\[.1cm]   
\nu_R^{(1)} \\ \vdots \end{array}\right) 
\end{equation} 
Here $m^{(0,0)}_L\equiv M^{(\nu)}$ are the Majorana masses for the zero
modes from eq.~(\ref{nu}). The Majorana masses $m^{(i,j)}_{L}$
and $m^{(i,j)}_{R}$ 
are obtained from  eq.~(\ref{nu}) by inserting the wave
functions of the relevant states. $m_{KK,i}$ are the KK masses 
of the excited states. Since the KK states have a greater overlap
with the Higgs, the corresponding Majorana masses are larger than those
involving zero modes. We consider again the case $k=0.01M_5$ and 
$\lambda_{\tau}^{(5)}=10g^{(5)}$ implying $c_{\tau,L}=0.69$.
To estimate the magnitude of the effect we restrict ourselves
to a single neutrino flavor. 
For the first excited state we find $m^{(0,1)}_L=l\cdot 8.8$ keV,
$m^{(1,1)}_L=l\cdot 3.5$ MeV and $m^{(1,1)}_R=l\cdot 20$ keV.
For higher excited states these masses approximately do not change.
If we would have used a delta function-like Higgs profile,
 $m^{(i,j)}_{R}$ would be zero  since the wave functions
of $\nu_R^{(i)}$ are odd and vanish at the boundaries.
Truncating the KK tower at the first excitation, we find
for the lowest state (i.e.~the zero mode)
\begin{eqnarray} \label{KK}
\nu_1&\approx& \nu_L^{(0)}+\frac{m_R^{(1,1)}m_L^{(0,1)}}{m_{KK,1}^2}\nu_L^{(1)}
-\frac{m_L^{(0,1)}}{m_{KK,1}}\nu_R^{(1)}\\
&\approx& \nu_L^{(0)}+ 1.5\cdot10^{-18}l^2\nu^{(1)}_L- 8\cdot10^{-10}l\nu^{(1)}_R. \nonumber
\end{eqnarray}
In the second line we used numbers from the previous
example. These tiny admixtures have only a negligible 
input on the properties of the zero modes. The tiny correction
to the mass of the zero mode, for instance, is given by
\begin{eqnarray}
\delta m_{\nu_1} &\approx& \frac{m_R^{(1,1)}(m_L^{(0,1)})^2}{m_{KK,1}^2}\\
                    &\approx&1.3\cdot10^{-14}l^3 {\rm~eV}. \nonumber
\end{eqnarray}
The contributions of the higher KK states are suppressed by larger KK
masses and will affect our result by at most a factor of order unity. 
In a similar way to (\ref{nu_mass}) the charged lepton masses (\ref{3.2a}) also 
couple zero modes and KK states. Here too the large KK masses
suppress corrections to charged lepton masses and mixings from 
the excited states. Thus, the KK states can be safely neglected 
in the discussion of neutrino mixings.

\section{Dirac vs. Majorana Neutrinos}
We finally discuss some experimental signatures to distinguish 
the presented scenario of Majorana neutrinos from the Dirac 
neutrino scheme discussed in refs.~\cite{GN,HS3}. 

\begin{figure}[t] 
\begin{picture}(100,160)
\put(95,-0){\epsfxsize7cm \epsffile{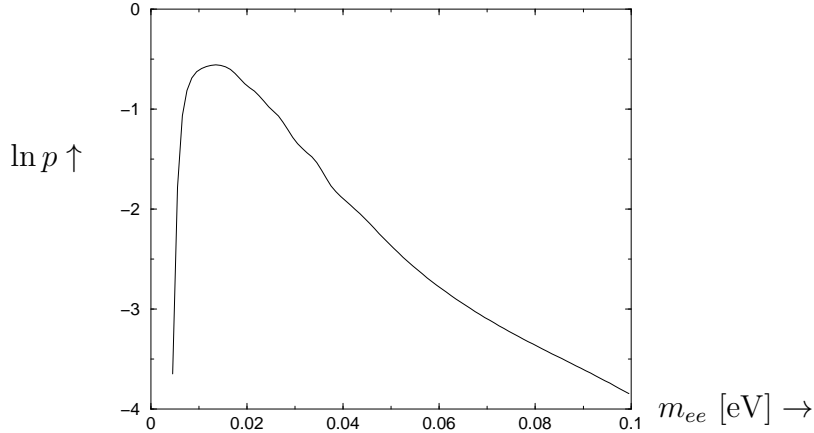}}
\put(55,100){{$\ln p\uparrow$}} 
\put(300,5){$m_{ee} ~[{\rm eV}]\rightarrow$}
\end{picture} 
\caption{Fraction of parameter sets $p$ which reproduce
the $\Delta m^2$'s as a function of the electron neutrino Majorana mass 
$m_{ee}$ for $c_L=0.72$ 
(in logarithmic scale).
}
\label{f_3}
\end{figure}

The most direct evidence
for a Majorana mass of neutrinos would be the discovery of
neutrinoless double beta decay ($0\nu\beta\beta$). 
The non-observation of this process in the Heidelberg-Moscow 
experiment implies an upper bound on 
$|M^{(\nu)}_{11}|\equiv m_{ee}<0.35$ eV \cite{HMC}.  There are
plans to bring this limit down to about 0.01 eV \cite{KK}.
If all neutrinos are at the same position in the extra dimension,
we expect $m_{ee}\sim0.02$ eV, which is far below the current
experimental sensitivity but within reach of future experiments. 
This result does not depend on what solution to the solar
neutrino anomaly is realized.  Fig.~\ref{f_3} illustrates the 
unnaturalness of a large Majorana mass in our scenario. As a
function of $m_{ee}$ we show in a
logarithmic scale the fraction of parameter sets $p$
which reproduce the $\Delta m^2$'s required for the solar LMA 
solution and for the atmospheric neutrino anomaly. (Including 
also the constraints from the mixing angles would reduce
the number of viable parameter sets by another order 
of magnitude.) We assume
all neutrinos to be at the same location with $c_L=0.72$, and take 
$1/2<|\tilde{l}_{ij}|<2$. To tune $m_{ee}$ we set $\tilde{l}_{11}=1$
and vary $\Lambda$. For small values of $m_{ee}$ (i.e.~small $\Lambda$)
it becomes impossible to accommodate the atmospheric $\Delta m^2$,
leading to $p=0$. For large Majorana masses $p$ becomes
exponentially small since it is highly improbable to accidentally
generate $\Delta m^2\ll m^2$.
The peak around $m_{ee}\approx0.015$ eV corresponds to the fraction 
of 28 \% quoted in table \ref{t_L}.
For larger ranges of $\tilde{l}_{ij}$ the distribution is spread out 
further.

Once the electron neutrino is localized closer towards the Planck-brane
$0\nu\beta\beta$ becomes drastically suppressed proportional to
$\epsilon^2$. For the clearly favored LMA scenario we find 
$m_{ee}\sim0.001$ eV, which is too small to be detected in the
near future. For the SMA, LOW and VAC solutions $m_{ee}$ is
even smaller. It is therefore questionable if our model induces
$0\nu\beta\beta$ at a detectable level. In any case the Majorana
masses we find are far below the recently claimed evidence
of about 0.4 eV \cite{KDHK}. 

Tritium endpoint (beta decay) experiments
are sensitive to neutrino masses of Majorana
and Dirac type.  The MAINZ collaboration plans 
to look for neutrino masses down to about 0.3 eV \cite{mainz},
which is ,however, still an order of magnitude larger
than the typical neutrino masses in the 
scenario considered.

In refs.~\cite{HS2,HS3} we imposed lepton number symmetry to make the
proton stable and eliminate Majorana neutrino masses from 
non-renormalizable operators. If in the current framework we 
want to keep Majorana masses but still forbid proton decay,
lepton parity is an attractive possibility. It still would allow
baryon number violating processes, such as neutron anti-neutron
oscillations, which might be close to the observational bound \cite{HS3}.
Without some symmetry dimension six operators inducing proton decay 
have to be multiplied by small couplings of order $10^{-8}$. 
The proton lifetime then should not be too far above the 
present experimental constraint. 
  
The scenario with Dirac fermions could manifest itself through
lepton flavor violating decays. For instance, the rate for 
$\mu\rightarrow e\gamma$ is considerably enhanced by 
the presence of sterile neutrino KK states, which spoil
the GIM cancellation of the SM. If the SM neutrinos are
confined to the TeV-brane, the branching ratio for  
$\mu\rightarrow e\gamma$ is above the experimental 
limit of about $10^{-11}$, unless the KK scale is above 
25 TeV \cite{K00}. In the case of bulk SM neutrinos   
$\mu\rightarrow e\gamma$  does not constrain the KK
scale anymore, but the branching ratio may still be
close to the experimental bound.

\section{Conclusions}
Neutrino masses generated by dimension five interactions in the SM are
of magnitude $\lsim10^{-5}$ eV, and consequently the
atmospheric neutrino anomaly remains problematic. 
In contrast, such masses are significantly larger in the 5D warped SM, and
we have shown that an explanation of the solar and atmospheric
neutrino anomalies can be realized, based on the idea of neutrino mass
anarchy. We also provide estimates for the mixing
angle $U_{e3}$  as well as the parameter $m_{ee}$ that appears in the amplitude for
neutrinoless double beta decay. No new particles such as SM singlet neutrinos
are needed to implement the scenario. Indeed, because of the warped geometry,
the usual see saw mechanism will not work in its simplest form, without 
either giving up the resolution of the gauge hierarchy problem, or invoking
supersymmetry.

\section*{Acknowledgements}
We thank David Emmanuel Costa, Holger Nielsen and  Chin-Aik Lee
for helpful discussions, and Wilfried Buchm\"uller for valuable comments 
on the draft.
This work was supported in part 
by DOE under contract DE-FG02-91ER40626. 
We thank the Alexander von Humboldt Stiftung for providing 
the impetus for this collaboration. Q.S.~also acknowledges
the hospitality of the Theory Group at DESY where this work
was completed.

\end{document}